# Spreading dynamics of a 2SIH2R, rumor spreading model in the homogeneous network

Yan Wang, [1,2] Feng Qing, [1,2] Jian-Ping Chai[1,2] and Ye-Peng Ni[1,2]

[1] State Key Laboratory of Media Convergence and Communication, Communication University of China, Beijing /100024, China.
[2] School of Data Science and Intelligent Media, Communication University of China, Beijing /100024, China.

Correspondence should be addressed to Feng Qing; qing329364213@cuc.edu.cn

## Abstract

In the era of the rapid development of the Internet, the threshold for information spreading has become lower. Most of the time, rumors, as a special kind of information, are harmful to society. And once the rumor appears, the truth will follow. Considering that the rumor and truth compete with each other like light and darkness in reality, in this paper, we study a rumor spreading model in the homogeneous network called *2SIH2R*, in which there are both spreader1(people who spread the rumor) and spreader2(people who spread the truth). In this model, we introduced discernible mechanism and confrontation mechanism to quantify the level of people's cognitive abilities and the competition between the rumor and truth. By mean-field equations, steady-state analysis and numerical simulations in a generated network which is closed and homogeneous, some significant results can be given: the higher discernible rate of the rumor, the smaller influence of the rumor; the stronger confrontation degree of the rumor, the smaller influence of the rumor; the large average degree of the network, the greater influence of the rumor but the shorter duration. The model and simulation results provide a quantitative reference for revealing and controlling the spread of the rumor.

## Introduction

With the continuous emergence of social media platforms, the traditional media era has gradually turned into the self-media era, and information dissemination has become faster, wider in scope, and deeper than ever [1]. Rumors, as a special kind of information, have greatly increased the possibility of artificial release of rumors due to their own confusion, timeliness and psychological satisfaction to the people who spread the rumor. Coupled with the self-media era, the threshold for spreading rumors is further lowered [2]. In today's society, there are some people who use people's public psychology to create rumors to obtain benefits from it [3-4]. This behavior will cause public panic and harm society. Therefore, in order to reveal the law of rumors dissemination and reduce the negative impact of rumors on society, it is necessary to establish a suitable mathematical model to analyse the characteristics and mechanisms of rumors dissemination process.

In the 1960s, Daley and Kendall [5] proposed the classic rumor spreading model, the DK model. The model divides the population into three categories: people who have never heard of rumors (Ignorant), people who spread rumors (Spreader), and people who have heard the rumors but do not spread (Stifler). The form to reflect the probability of an individual to switch between these three categories in order to characterize the individual reaction after receiving a rumor. Zanette [6-7] used complex network theories to study the spread of rumors. He established a rumor spreading model in small-world networks and proved the threshold of rumor spreading. Moreno [4] established a SIR (susceptible-infective-refractory) model in a scale-free network and analysed the simulation results. In the process of research, some scholars adjusted the SIR model according to the research purpose, and applied it to the complex network getting numerous great consequences. Wan [8] studied the propagation process of the adjusted SIR model in a homogeneous network, and proposed two strategies for network rumor immunity: active immunity and passive immunity. Zhao [9] introduced a media report mechanism and studied the influence of media reports on rumors. Askarizadeh [10] introduced an anti-rumor mechanism and proposed a game model to analyse the process of rumor spread in social networks, and concluded that anti-rumor will affect the spread of rumors. Zhang [11] established I2S2R dynamic rumor propagation models in homogeneous and heterogeneous networks. Huo [12] proposed that the SIbInIu model divides the population into four categories, and concluded that the losing-interest rate and stifling rate have a negative impact on the scale of the final spread of rumors. Deng [13] introduced the forgetting and memory mechanisms in the process of studying the spread of rumors. Gu [14] established a SEIR model on Facebook's user data set, and concluded that acquaintance immunization is the best solution to curb online rumors through the comparison of multiple immunization strategies. Wang [15] established the SIRaRu model and proposed that when the ignorant comes into contact with the spreader, the ignorant believes the rumor or not with probability, corresponding to the formation of two stifle states. In 2014, Wang [16] also considered the existence of multiple rumors in a network, and one type of rumors would be affected by another type of rumors. Wang [17], Yang [18], and Xia [19] introduced the hesitation mechanism in their respective models. People who heard the rumors temporarily did not spread, when they were heard the rumors again, they became the spreaders with probability. Based on the DK model, Huo [20] divided spreaders into two categories, namely, spreaders with high activity and spreaders with low activity. Later, Huo [21] introduced the indiscernible degree mechanism in the model to describe the individual's cognitive ability. Ran [22] introduced a rumor rejection mechanism while considering the impact of individual differences on the spread of rumors, and established an IWSR rumor spread model. Dong [23] proposed a double-identity rumor spreading model, that is, in addition to Ignorant, Spreader, and Stifler, each network node also has one of three other identities, namely, rumor creator, rumor controller and normal user. Zan [24] established the SICR model to introduce the counterattack mechanism of rumors, that is, when the spreader contacts the counterattack, the spreader becomes a stifler with probability.

In the above studies, many have made great contributions to the theoretical research on the process of rumor spreading on complex networks. However, there are two shortcomings in the theory to need to be improved. The first one is that, in reality, the discernible degree of the rumor is an important variable, but most previous studies did not quantify this. Allport and Postman [25] believe that there are three conditions for the generation and spread of rumors: the first one is the lack of information; the second one is people's anxiety; the third one is that the society is in crisis. Based on this, they proposed a classical formula: $rumors = i \times a$ (where $i$ represents the importance of information, and $a$ represents the degree of unknowability of the event). The other improvement to be made is that no research has been done on the spread of truth, the opposite of rumors. With the rumor, there is also the truth. In reality, there are always some wise men who can reveal the rumor and spread the truth, in which time there will be a confrontation relationship between the rumors and the truth [24]. Based on this, we divide the population into six categories: people who have never heard of rumors or truth (ignorant), people who spread rumors (spreader1), people who spread truth (spreader2), people who have heard the rumors but do not spread temporarily (hesitant1), people who have heard the rumors but do not spread(stifler1), people who have heard the truth but do not spread(stifler2), and propose the *2SIH2R* model with the discernible mechanism and the confrontation mechanism.

The organization of the paper is the following. In Section 2, the *2SIH2R* model is defined, and the mean-field equations of the model are established in the homogeneous network [16,26]. In Section 3, we study the rumor spreading threshold of model propagation by changing initial conditions and parameters, and extend the spreading threshold under special circumstances [27] to general conditions. In Section 4, through simulation, we study the influence of discernible mechanism, confrontation mechanism, and average degree on the rumor. In Section 5, conclusions of the paper and future work are given.

## 2SIH2R Rumor Spreading Model

We consider a closed and mixed population composing of *N* individuals as a complex network, where individuals and their contacts can be represented by vertexes and edges. This network can be described by an undirected graph $G = (V, E)$ where *V* denotes the vertexes and *E* represents the edges. At each time $t$, the people in the network can be divided into: $S_1$, $S_2$, $I$, $H$, $R_1$, $R_2$, separately, representing for people who spread the rumor, people who spread truth, people who have never heard of the rumor or truth, people who have heard the rumor but do not spread temporarily, people who have heard the rumor but do not spread, people who have heard the truth but do not spread. The rumor spreading process of the *2SIH2R* can be seen in Figure 1.

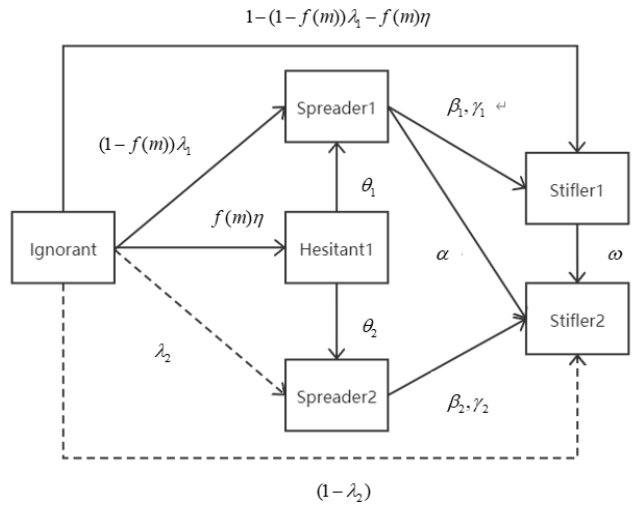

Figure 1 *2SIH2R* rumor spreading process

In Figure 1, the solid/dotted line from the "Ignorant" represents that the ignorant contact with spreader1/spreader2, and the rumor spreading rules of the *2SIH2R* model can be summarized as follows.

(1) We use $m$ to describe the discernible rate of the rumor, and $f(m)$ to describe people's ability to reveal rumors. The function of $f$ is to map the characteristics of the rumor to the characteristics of the people. The greater the $m$, the greater probability that the rumor will be revealed. The greater the $f(m)$, the greater probability that the people will not believe the rumor immediately. We assume that there is a positive correlation between $m$ and $f(m)$.

(2) When an ignorant encounters a spreader1, there are three possible outcomes: (i) the ignorant may believe the rumor and spread it with probability $(1-f(m))\lambda_1$. The $\lambda_1$ namely is rumor spreading rate; (ii) the ignorant may not believe the rumor immediately and hesitate to spread it with probability $f(m)\eta$. The $\eta$ namely is potential spreading rate; (iii) the ignorant may have no response to the rumor with probability $1-(1-f(m))\lambda_1 - f(m)\eta$, because of having no interest of the rumor.

(3) When an ignorant encounters a spreader2, there are two possible outcomes: the ignorant may believe the truth and spread it with probability $\lambda_2$, called truth spreading rate; (ii) the ignorant may have no response to the truth with probability $(1-\lambda_2)$, because of having no interest of the truth. Since the truth is generally issued by an authoritative organization, or there is evidence to support it, the truth is relatively more objective, accurate, and clear. It is easier for the ignorant to judge, and not easy to become a hesitant.

(4) We consider that the hesitant1 have desire to spread the information, because of the suspicion of the rumor and the environmental impact when they receive the rumor, they didn't spread the rumor immediately. In the time of hesitation, hesitant1 may believe the rumor to spread it, or they may discover the truth and spread it. So, we assume that at each step, the hesitant1 will spontaneously become people who spread the rumor(spreader1) with probability $\theta_1$, and people who spread truth(spreader2) with probability $\theta_2$.

(5) When a spreader1(spreader2) encounters another spreader1(spreader2), he/she could think the rumor(truth) is widely known. So, the spreader1(spreader2) may lose spreading enthusiasm and become a stifler1(stifler2) with rumor(truth) losing-interest rate $\beta_1$ ($\beta_2$).

(6) At each step, spreader1(spreader2) becomes a stifler1(stifler2) spontaneously with probability $\gamma_1(\gamma_2)$, called rumor(truth) forgetting-rate.

(7) At each step, the stifler1 will spontaneously become stifler2 with probability $\omega$, because of the improvement of their own cognitive level.

(8) When a spreader1 encounters spreader2, the spreader1 will believe the truth rather than the rumor with probability $\alpha$, because of the confrontation mechanism between the truth and rumor. The $\alpha$ namely is confrontation rate.

Moreover, the *2SIH2R* model is applied to a generated network which is a closed and homogeneous population consisting of *N* individuals [17,28]. We use $S_1(t)$, $S_2(t)$, $I(t)$, $H(t)$, $R_1(t)$, $R_2(t)$, separately, to represent the densities of spreader1, spreader2, ignorant, hesitant1, stifler1, stifler2, and at any step, we have the normalization condition:

$$I(t) + S_1(t) + S_2(t) + H(t) + R_1(t) + R_2(t) = 1$$

According to the rumor spreading rules, the mean-field equation of *2SIH2R* model can be expressed as follows:

$$\frac{dI(t)}{dt} = -<k>(S_1(t) + S_2(t))I(t) \tag{1}$$

$$\frac{dS_1(t)}{dt} = (1-f(m))\lambda_1 <k> S_1(t)I(t) + \theta_1 H(t) - \alpha <k> S_1(t)S_2(t)$$

$$-\beta_1 <k> S_1(t)(S_1(t)+R_1(t)+H(t)) - \gamma_1 S_1(t) \qquad (2)$$

$$\frac{dS_2(t)}{dt} = \lambda_2 <k> S_2(t)I(t) + \theta_2 H(t) - \beta_2 <k> S_2(t)(S_2(t)+R_2(t)) - \gamma_2 S_2(t) \qquad (3)$$

$$\frac{dH(t)}{dt} = f(m)\eta <k> S_1(t)I(t) - (\theta_1+\theta_2)H(t) \qquad (4)$$

$$\frac{dR_1(t)}{dt} = (1-(1-f(m))\lambda_1 - f(m)\eta) <k> S_1(t)I(t)$$

$$+\beta_1 <k> S_1(t)(S_1(t)+R_1(t)+H(t)) + \gamma_1 S_1(t) - \omega R(t) \qquad (5)$$

$$\frac{dR_2(t)}{dt} = (1-\lambda_2) <k> S_2(t)I(t) + \alpha <k> S_1(t)S_2(t)$$

$$+\beta_2 <k> S_2(t)(S_2(t)+R_2(t)) + \gamma_2 S_2(t) + \omega R_1(t) \qquad (6)$$

Where $<k>$ represents the average degree of the generated network.

## Steady-state Analysis

In this section, we will consider the three situations of the model. When the system reaches the steady state, there is neither spreader1 nor spreader2. So, we can give the condition in the final state: $S_1 = \lim_{t\to\infty} S_1(t) = 0$, $S_2 = \lim_{t\to\infty} S_2(t) = 0$, $H = \lim_{t\to\infty} H(t) = 0$, and $\lim_{t\to\infty}(I(t)+R_1(t)+R_2(t)) = 1$. It is assumed that $I = \lim_{t\to\infty} I(t)$, $R_1 = \lim_{t\to\infty} R_1(t)$, $R_2 = \lim_{t\to\infty} R_2(t)$. The final size of the rumor(truth) $R_1$ ($R_2$) will be calculated to measure the level of the rumor(truth) influence[12], and $R = R_1 + R_2$ is used to measure the level of influence of the model. We will study the rumor spreading threshold of the model by analysing the final size of $R$. The sum of Eq.(5) and Eq.(6) is divided by Eq.(1), we have:

$$\frac{dR(t)}{dI(t)} = \frac{d(R_1(t)+R_2(t))}{dI(t)}$$

$$= -\frac{(1-(1-f(m))\lambda_1 - f(m)\eta)S_1(t) + (1-\lambda_2)S_2(t)}{S_1(t)+S_2(t)}$$

$$-\frac{\beta_1 S_1(t)(S_1(t)+R_1(t)+H(t)) + \beta_2 S_2(t)(S_2(t)+R_2(t)) + \alpha S_1(t)S_2(t)}{(S_1(t)+S_2(t))I(t)}$$

$$-\frac{\gamma_1 S_1(t) + \gamma_2 S_2(t)}{<k>(S_1(t)+S_2(t))I(t)} \qquad (7)$$

## Steady-state Analysis of Rumor

At the beginning of model spreading, in this situation, we assumed that there is only one spreader1 who spreads the rumor, and there is no truth. So, the initial condition can be given: $S_2(0) = 0$, $\omega = 0$, $S_1(0) = \frac{1}{N} \approx 0$, $I(0) = \frac{N-1}{N} \approx 1$, $H(0) = R_1(0) = R_2(0) = 0$. After a while the number of spreader1 will increase to the top, then it reduces to zero at which time the system reaches stability.

Since $S_2(0) = 0$, $\omega = 0$, Eq.(3) and Eq.(6), we can know $S_2(t) = 0$ and $R_2(t) = 0$. So, there is the normalization condition $I(t) + S_1(t) + H(t) + R_1(t) = 1$. Considering the above-mentioned conditions, Eq.(7) becomes

$$\frac{dR(t)}{dI(t)} = \frac{d(R_1(t) + R_2(t))}{dI(t)}$$

$$= -(1 - (1 - f(m))\lambda_1 - f(m)\eta) - \frac{\beta_1(1 - I(t))}{I(t)} - \frac{\gamma_1}{<k>I(t)}$$

$$= -c - \frac{\beta_1(1 - I(t))}{I(t)} - \frac{\gamma_1}{<k>I(t)}$$

$$\Rightarrow dR(t) = -cdI(t) - \frac{\beta_1(1-I(t))}{I(t)}dI(t) - \frac{\gamma_1}{<k>I(t)}dI(t)$$

$$\Rightarrow \int_0^\infty dR(t)dt = \int_0^\infty (-cdI(t) - \frac{\beta_1(1-I(t))}{I(t)}dI(t) - \frac{\gamma_1}{<k>I(t)}dI(t))dt$$

$$\Rightarrow R = (\beta_1 - c)(I - 1) - (\beta_1 + \frac{\gamma_1}{<k>})\ln I$$

$$\Rightarrow (\beta_1 - c + 1)R = -(\beta_1 + \frac{\gamma_1}{<k>})\ln(1-R)$$

$$\Rightarrow \frac{\beta_1 - c + 1}{-(\beta_1 + \frac{\gamma_1}{<k>})} R = \ln(1-R)$$

$$\Rightarrow R = 1 - e^{-\frac{\beta_1 - c + 1}{\beta_1 + \frac{\gamma_1}{<k>}}R}$$

$$\Rightarrow R = 1 - e^{-\varepsilon R} \qquad (8)$$

Where $c = 1 - (1 - f(m))\lambda_1 - f(m)\eta$ and $\varepsilon = \frac{\beta_1 - c + 1}{\beta_1 + \frac{\gamma_1}{<k>}}$. Only when $\varepsilon > 1$, will get Eq.(8) a non-zero solution. For $f(m) \neq 1$, the following condition will be satisfied:

$$\lambda_{1c} = \frac{\gamma_1 - <k>f(m)\eta}{<k>(1 - f(m))} \qquad (9)$$

Due to the confrontation mechanism, when $0 \leq \lambda_1 \leq \lambda_{1c}$, the rumor must not spread widely in the generated network.

**Steady-state Analysis of Truth**

It is assumed that the government or authoritative media have already begun to spread the truth before a rumor event occurs. Then when a rumor event occurs, there will be no rumor spreader in the population. So, in this situation, the initial condition can be given: $S_1(0) = 0$. Then we can prove $S_1(t) = 0$ from Eq.(3), and $R_1(t) = 0$ from Eq.(5). Since there are no hesitant in the network at beginning, we can prove $H(0) = 0$ from Eq.(4). Next, we can follow the proof process in the previous part to get:

$$\lambda_{2c} = \frac{\gamma_2}{<k>} \tag{10}$$

So, when $0 \leq \lambda_1 \leq \lambda_{1c}$, the rumor must not spread widely in the generated network.

**Steady-state Analysis of 2SIH2R model**

In this part, we consider a relatively general situation. At the beginning of model spreading, in this situation, we assumed that there is one spreader1 who spreads the rumor, and one spreader2 who spreads the truth. So, the initial conditions can be given: $I(0) = \frac{N-2}{N} \approx 1$, $S_1(0) = S_2(0) = \frac{1}{N} \approx 0$, $H(0) = R_1(0) = R_2(0) = 0$.

Moreover, it is worth noting that in this situation, due to the complicated spreading process, we can not follow the proof process in the previous part to get the condition of *2SIH2R* model spreading threshold. Therefore, this part re-starts from the initial conditions and gets the condition of *2SIH2R* model spreading threshold.

We use $i(t), s_1(t), s_2(t), h(t), r_1(t), r_2(t)$, separately, to represent from Eq.(1) to Eq.(6), and from the initial conditions we can know:

$$i(0) = -2 <k> \frac{(N-2)}{N^2} \tag{11}$$

$$s_1(0) = (1 - f(m))\lambda_1 <k> \frac{N-2}{N^2} - (\alpha + \beta_1) <k> \frac{1}{N^2} - \frac{\gamma_1}{N} \tag{12}$$

$$s_2(0) = \lambda_2 <k> \frac{N-2}{N^2} - \beta_2 <k> \frac{1}{N^2} - \frac{\gamma_2}{N} \tag{13}$$

$$h(0) = f(m)\eta <k> \frac{N-2}{N^2} \tag{14}$$

$$r_1(0) = (1-(1-f(m))\lambda_1 - f(m)\eta) <k> \frac{N-2}{N^2} + \beta_1 <k> \frac{1}{N^2} + \frac{\gamma_1}{N} \tag{15}$$

$$r_2(0) = (1-\lambda_2) <k> \frac{N-2}{N^2} + \alpha <k> \frac{1}{N^2} + \beta_2 <k> \frac{1}{N^2} + \frac{\gamma_2}{N} \tag{16}$$

Here $i(0)$ represents the instantaneous rate of change of the ignorant when $t=0$. The $r(0) = r_1(0) + r_2(0)$ represents the instantaneous rate of change of the stifler1 and stifler2. If the *2SIH2R* model can work successfully, at $t=0$, some ignorant people must become the other five categories. So, the number of ignorant people decreases, and the number of the other five categories increases, which means $i(0)<0$, $|i(0)|>|r(0)|$ and $r(0) \geq 0$.

From Eq.(11) to Eq.(16) we have

$$i(0) + s_1(0) + s_2(0) + h(0) + r_1(0) + r_2(0) = 0 \tag{17}$$

Since $r(0) = r_1(0) + r_2(0) \geq 0$, we have

$$s_1(0) + s_2(0) + h(0) > 0 \tag{18}$$

So,

$$\begin{aligned} s_1(0)+s_2(0)+h(0) = & (1-f(m))\lambda_1 <k> \frac{N-2}{N^2} - (\alpha+\beta_1) <k> \frac{1}{N^2} - \frac{\gamma_1}{N} \\ & +\lambda_2 <k> \frac{N-2}{N^2} - \beta_2 <k> \frac{1}{N^2} - \frac{\gamma_2}{N} + f(m)\eta <k> \frac{N-2}{N^2} > 0 \end{aligned} \tag{19}$$

From Eq.(19) we have:

$$(1-f(m))\lambda_1 + \lambda_2 + f(m)\eta > \frac{\alpha+\beta_1+\beta_2}{N-2} + \frac{N(\gamma_1+\gamma_2)}{(N-2)<k>} \tag{20}$$

When $N \to \infty$, the following result can be obtained:

$$(1-f(m))\lambda_1 + \lambda_2 > \frac{\gamma_1}{<k>} - f(m)\eta + \frac{\gamma_2}{<k>} \tag{21}$$

So, if the rumor and the truth can spread widely in the generated network which is closed and homogeneous, the $\lambda_1$ and $\lambda_2$ should be satisfied the Eq.(21). Next, set $\lambda_2=0$, $\gamma_2=0$, we can get the Eq.(9). And by setting $\lambda_1=0$, $f(m)=0$, we can get the Eq.(10). So, we can conclude that the third general situation contains the first two special situations.

## Numerical simulation

In this section, through numerical simulation, we study the influence of discernible mechanism, confrontation mechanism, and average degree on the rumor. According to the *2SIH2R* Rumor Spreading Model and existing research results [29-31], we perform numerical simulation in a generated homogeneous network, where $<k>=8$, $N=10^5$. It is assumed that there are one spreader1 and one spreader2 at the time $t=0$. So, $I(0) = \frac{N-2}{N} \approx 1$, $S_1(0) = S_2(0) = \frac{1}{N} \approx 0$, $H(0) = R_1(0) = R_2(0) = 0$.

Figure 2 displays the change of density of six categories (spreader1, spreader2, stifler1, stifler2, ignorant, hesitant) over time with $f(m)=0.7m$, $m=0.3$, $\lambda_1=\lambda_2=0.7$, $\eta=0.8$, $\theta_1=0.5$, $\theta_2=0.3$, $\beta_1=\beta_2=0.3$, $\gamma_1=\gamma_2=0.1$, $\omega=0$, $\alpha=0.5$. Unless otherwise specified, the above parameters are used in this section. It can be seen from Figure 2 that the density of the ignorant decreases rapidly and the other 5 categories increases to their peak, separately in a short time. As the model spreads further, the densities of spreader1 and spreader2 will continue decreasing until it reaches zero, which means the *2SIH2R* model gets into the steady state.

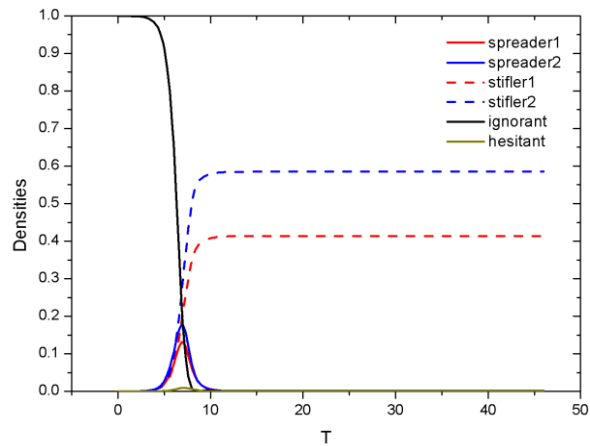

Figure 2 Density of 6 categories of people in homogeneous network

Figure 3 displays the change of density of spreader1, under the change of parameter $m$. It can be seen that the greater $m$ (the bigger $f(m)$), the stronger discernible mechanism, the smaller impact of the rumor, because of the decreasing peak. At the same time, from the Figure 4, as $m$ increases, the final size of stifler1 also decreases. But the time to peak of spreader1 and stifler1 has not changed significantly. In Figure 5, it also can be seen that the final size of stifler1 decreases with increasing $f(m)$, but the stifler2 increases. In summary, as $m$ increases, the instantaneous maximum influence and the final influence range of the rumor will decrease but the truth increase.

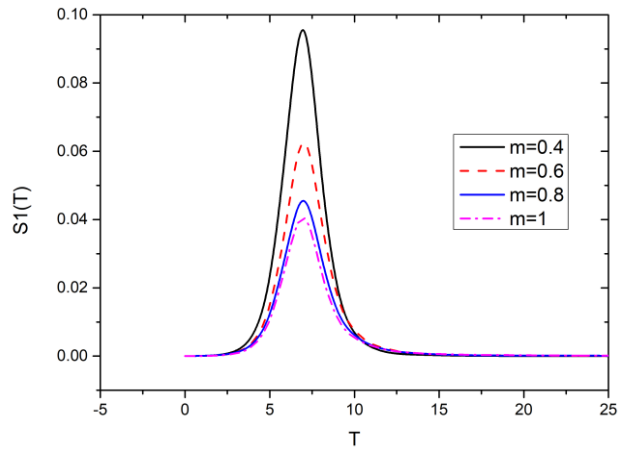

Figure 3 Density of spreader1 over time for different values of $m$

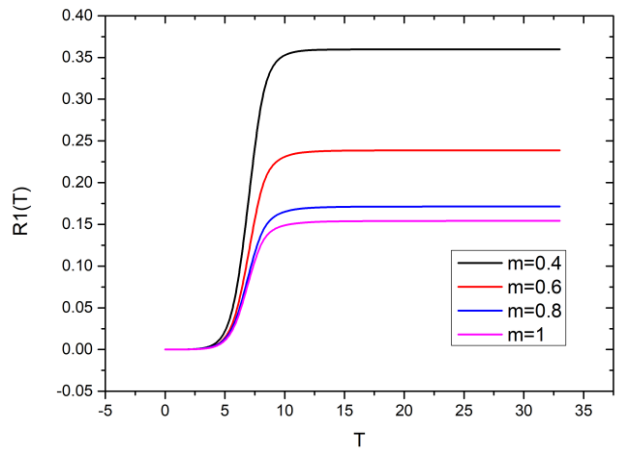

Figure 4 Density of stifler1 over time for different values of $m$

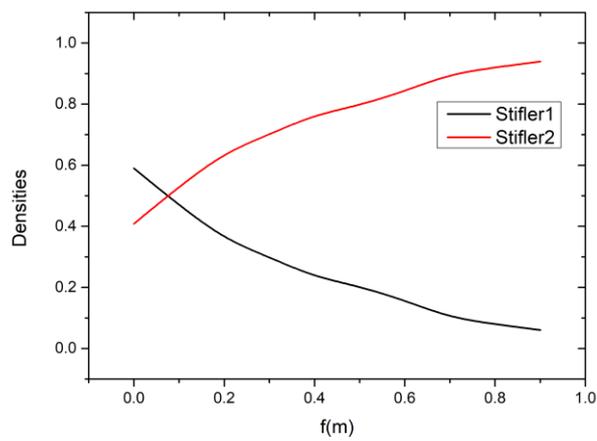

Figure 5 Density of stifler1 and stifler2 for different values of $f(m)$ at steady-state

Figure 6 displays the change of density of spreader1, under the change of parameter $\alpha$. It can be seen that the greater $\alpha$, the stronger confrontation mechanism, the smaller impact of the rumor, because of the decreasing peak. At the same time, from Figure 7, as $\alpha$ increases, the final size of stifler2 increases, because some spreader1 change into stifler2 by the confrontation mechanism. In Figure 8, it can also be seen that the final size of stifler1 decreases with increased $\alpha$, but the stifler2 increases. In summary, as $\alpha$ increases, the instantaneous maximum influence and the final influence range of the rumor will decrease, while the final influence range of the truth will increase.

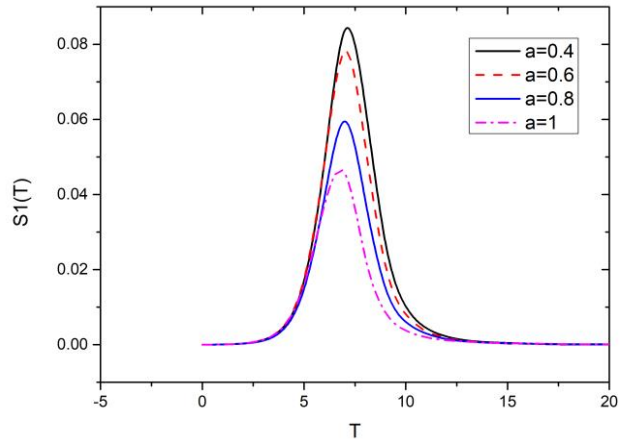

Figure 6 Density of spreader1 over time for different values of $\alpha$

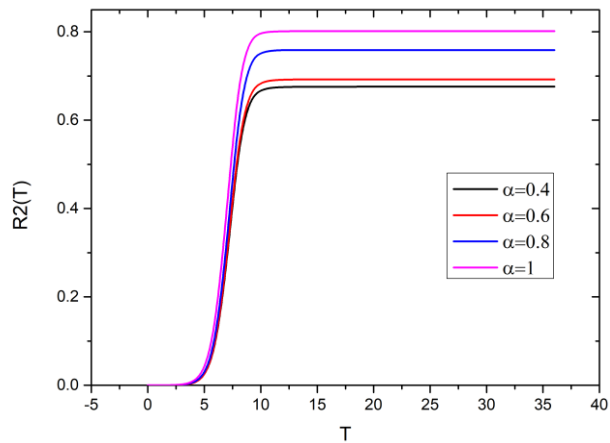

Figure 7 Density of stifler2 over time for different values of $\alpha$

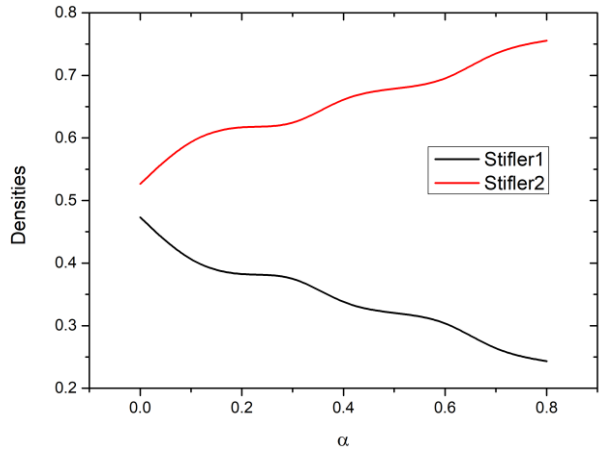

Figure 8 Density of stifler1 and stifler2 for different values of $\alpha$ at steady-state

Figure 9 displays the change of density of spreader1, under the change of parameter $<k>$. It can be seen that the greater $<k>$, the more people can be contacted by spreader1, the greater impact of the rumor, because of the increased peak and the shortened time of reaching the peak. Moreover, we can find that with increased $<k>$, the shape of the solid line becomes wider, which means that the duration of the rumor event is decreasing. In summary, as $<k>$ increases, the velocity and the range of the rumor spreading will increase, which means the influence of the rumor will increase significantly. But the duration of the rumor event will decrease.

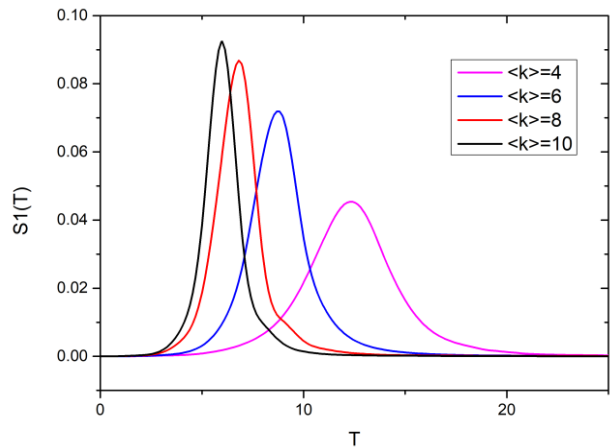

Figure 9 Density of spreader1 over time for different values of $<k>$

Figure 10 displays the change of density of stifler1 and stifler2, under the change of parameter $\omega$. It can be seen that the change of $\omega$ causes a huge impact on the rumor. As long as $\omega$ changes from 0 to 0.1, almost only stifler2 exist in the network when it reaches a steady-state, which means the rumors will not have a significant impact on us. So, this paper mainly studies the situation where $\omega=0$.

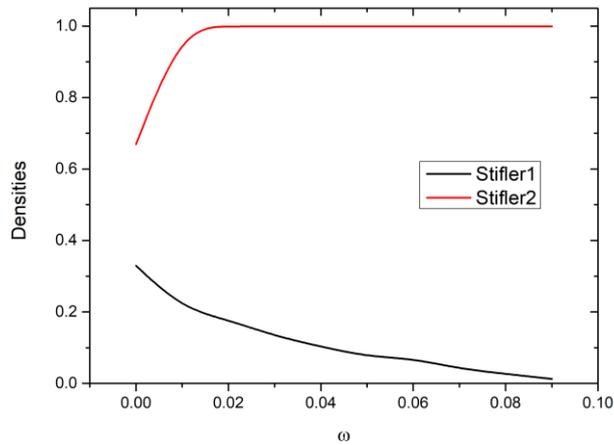

Figure 10 Density of stifler1 and stifler2 for different values of $\omega$ at steady-state

Figure 11 and Figure 12 display the final size $R$ ( which is the sum densities of stifler1 and stifler2 at steady-state) with $\lambda_1$ and $\lambda_2$. The redder the color is, the greater the value of $R$. In Figure 12, under the parameter $f(m)=0.5$, $\eta=0.1$, $\gamma_1=\gamma_2=0.8$, the spreading threshold condition can be distinguished roughly by the shade of color (as the black solid line denoted in the figure) which is basically consistent with the steady-state analysis from the previous section (as the black dash line denoted in the figure).

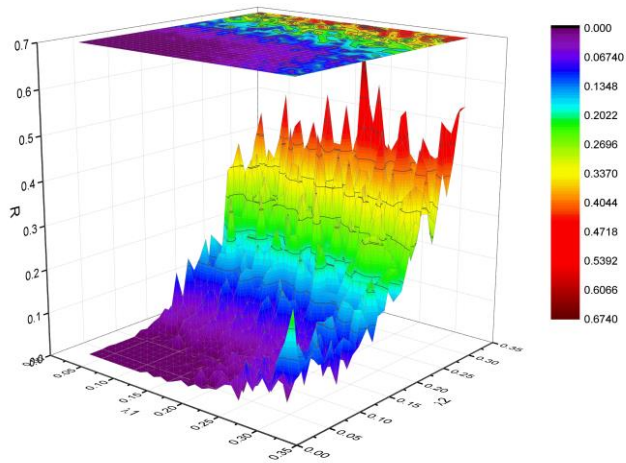

Figure 11 Density of $R$ with $\lambda_1$ and $\lambda_2$ under the parameter $f(m)=0.5$, $\eta=0.1$, $\gamma_1=\gamma_2=0.8$

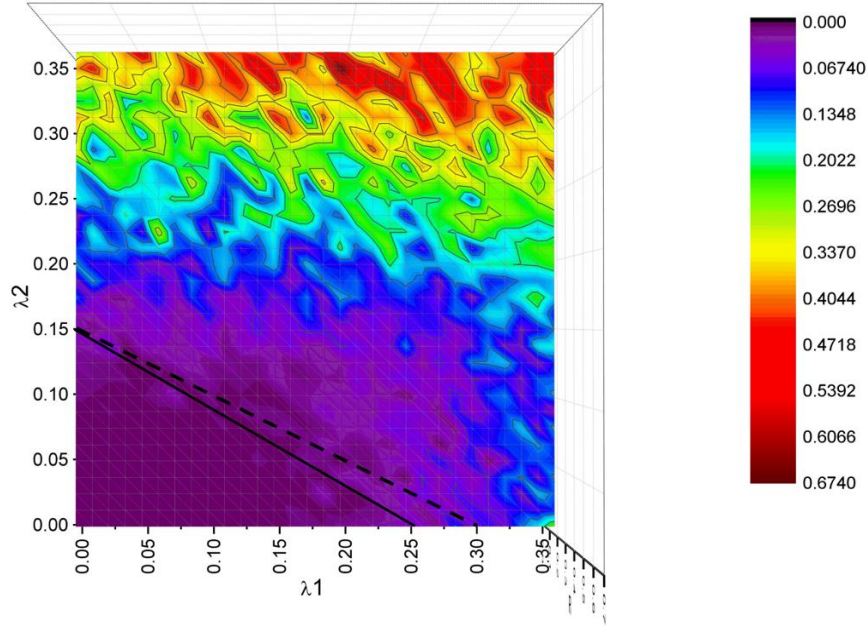

Figure 12 The vertical view of Figure 11

## Conclusions

Rumors as a kind of harmful information in most situation may cause large public panic. It is necessary to establish a suitable mathematical model to analyse the characteristics and mechanisms of the rumor dissemination process. In this paper, we propose a *2SIH2R* rumor spreading model in a generated homogeneous network, and some significant results can be given:

(1) We assumed that when an ignorant encounters a spreader1, the ignorant may change into hesitant1 because of the discernible mechanism, and when a spreader1 encounters a spreader2, the spreader1 may change into stifler2 because of the confrontation mechanism.
(2) Through changing parameters, the model can be simplified to the traditional SIR model and the SEIR model. From this point, the traditional model has been improved and the *2SIH2R* model is more universal.
(3) After establishing the mean-field equations of the *2SIH2R* model, we give the condition of *2SIH2R* model spreading threshold in three situations separately. When there is only one spreader1 at $t=0$ in the networks, the spreading threshold is $\lambda_{1c} = \frac{\gamma_1 - <k>f(m)\eta}{<k>(1-f(m))}$. When there is only one spreader2 at $t=0$ in the networks, the spreading threshold is $\lambda_{2c} = \frac{\gamma_2}{<k>}$. When there is one spreader1 and one spreader2 at $t=0$, the condition of *2SIH2R* model spreading threshold is $(1-f(m))\lambda_1 + \lambda_2 > \frac{\gamma_1 + \gamma_2}{<k>} - f(m)\eta$. When the condition is not satisfied, the rumor or the truth cannot spread widely in the crowd.
(4) From the numerical simulations, we can know that the higher the discernible degree is, the smaller influence of the rumor will be; The higher the confrontation rate is, the smaller influence of the rumor will be; The bigger the average degree is, the greater influence of the rumor will be, but the shorter duration is.

In the future, a further study of *2SIH2R* rumor spreading model will be conducted in the heterogeneous and some real networks. In this paper we assume that the social network is homogeneous, but in reality, lots of social networks have a more complex structure. And also, the real data may be analysed, because there are many subjective assumptions in the model and parameters setting process. The significance of the model can be better demonstrated through real data.

## Data Availability

The generated data used to support the findings of this study have not been made available because the data is randomly generated according to the rules.

## Conflicts of Interest

The author(s) declare(s) that there is no conflict of interest regarding the publication of this paper.

## Funding Statement

This paper is financially supported by Beijing Municipal Natural Science Foundation (9202018), Fundamental Research Funds for the Central Universities (CUC2000004, CUC19ZD002, 2018CUCTJ047), Academic Research Project of China Federation of Radio and Television Association(2020ZGLH015)

## References

[1] Zhang Ju -Ping, Guo Hao -Ming, Jing Wen -Jun and Jin Zhen, "Dynamic analysis of rumor propagation model based on true information spreader," *Acta Physica Sinica*, vol. (68), no. (15), pp. (193-204), 2019.

[2] Zhu Lin -He and Li Ling, "Dynamic analysis of rumor-spread-delaying model based on rumor-refuting mechanism," *Acta Physica Sinica*, vol. (69), no. (02), pp. (67-77), 2020.

[3] Amankwah-Amoah, Joseph, Antwi-Agyei et al., "Integrating the dark side of competition into explanations of business failures: Evidence from a developing economy," *European Management Review*, vol. (15), no. (1), pp. (97-109), 2017.

[4] Moreno Y, Nekovee M and Pacheco AF, "Dynamics of rumor spreading in complex networks," *Physical Review E*, vol. (69), no. (6), Article ID 066130, 2004.

[5] Daley D J and Kendall D G, "Epidemics and Rumours," *Nature*, vol. (204), no. (1118), 1964.

[6] Damián H Zanette, "Critical behavior of propagation on small-world networks," *Physical Review E*, vol. (64), Article ID 050901, 2001.

[7] Damián H Zanette, "Dynamics of rumor propagation on small-world networks," *Physical Review E*, vol. (65), no. (4), Article ID 041908, 2002.

[8] Wan Yi-Ping, Zhang Dong-Ge and Ren Qing-Hui, "Propagation and inhibition of online rumor with considering rumor elimination process," *Acta Physica Sinica*, vol. (64), no. (24), Article ID 240501, 2015.

[9]     Zhao Min, Chen Wenxia and Song Qiankun, "Research on a Rumor Spreading Model with Media Coverage," *Applied Mathematics and Mechanics*, vol. (39), no. (12), pp. (1400-1409), 2018.

[10]    Mojgan Askarizadeha, Behrouz Tork Ladania and Mohammad Hossein Manshaei, "An evolutionary game model for analysis of rumor propagation and control in social networks," *Physica A: Statistical Mechanics and its Applications*, vol. (523), no. (1), pp. (21-39), 2019.

[11]    Yuhuai Zhang and Jianjun Zhu, "Stability analysis of I2S2R rumor spreading model in complex networks," *Physica A: Statistical Mechanics and its Applications*, vol. (503), no. (1), pp. (862-881), 2018.

[12]    Liang'an Huo, Fan Ding and Yingying Cheng, "Dynamic analysis of a SIbInIu, rumor spreading model in complex social network," *Physica A: Statistical Mechanics and its Applications*, vol. (523), no. (1), pp. (924-932), 2019.

[13]    Shengfeng Deng and Wei Li, "Spreading dynamics of forget-remember mechanism," *Physical Review E*, vol. (95), no. (4), Article ID 042306, 2017.

[14]    Gu Yi-Ran and Xia Ling-Ling, "Propagation and inhibition of rumors in online social network," *Acta Physica Sinica*, vol. (61), no. (23), Article ID 238701, 2012.

[15]    Jiajia Wang, Laijun Zhao and Rongbing Huang, "SIRaRu rumor spreading model in complex networks," *Physica A: Statistical Mechanics and its Applications*, vol. (398), no. (15), pp. (43-55), 2014.

[16]    Jiajia Wang, Laijun Zhao and Rongbing Huang, "2SI2R rumor spreading model in homogeneous networks," *Physica A: Statistical Mechanics and its Applications*, vol. (431), no. (1), pp. (153-161), 2014.

[17]    Tao Wang, Juanjuan He and Xiaoxia Wang, "An information spreading model based on online social networks," *Physica A: Statistical Mechanics and its Applications*, vol. (490), no. (15), pp. (488-496), 2018.

[18]    Anzhi Yang, Xianying Huang, Xiumei Cai, Xiaofei Zhu and Ling Lu, "ILSR rumor spreading model with degree in complex network," *Physica A: Statistical Mechanics and its Applications*, vol. (531), no. (1), Article ID 121807, 2019.

[19]    Ling-Ling Xia, Guo-Ping Jiang, Bo Song and Yu-Rong Song, "Rumor spreading model considering hesitating mechanism in complex social networks," *Physica A: Statistical Mechanics and its Applications*, vol. (437), no. (1), pp. (295-303), 2015.

[20]    Liang'an Huo, Li Wang, Naixiang Song, Chenyang Ma and Bing He, "Rumor spreading model considering the activity of spreaders in the homogeneous network," *Physica A: Statistical Mechanics and its Applications*, vol. (468), no. (15), pp. (855-865), 2017.

[21]    Liang'an Huo and Yingying Cheng, "Dynamical analysis of a IWSR rumor spreading model with considering the self-growth mechanism and indiscernible degree," *Physica A: Statistical Mechanics and its Applications*, vol. (536), no. (15), Article ID 120940, 2019.

[22]    Ran Maojie, Liu Chao, Huang Xianying et al., "Rumor spread model considering difference of individual interest degree and refutation mechanism," *Journal of Computer Applications*, vol. (38), no. (11), pp. (3312-3318), 2018.

[23]    Xuefan Dong, Yijun Liu, Chao Wu, Ying Lian, and Daisheng Tang, "A double-identity rumor spreading model," *Physica A: Statistical Mechanics and its Applications*, vol. (528), no. (15), Article ID 121479, 2019.

[24] Yongli Zan, Jianliang Wu, Ping Li and Qinglin Yu, "SICR rumor spreading model in complex networks: Counterattack and self-resistance," *Physica A: Statistical Mechanics and its Applications*, vol. (405), no. (1), pp. (159-170), 2014.

[25] G.W. Allport, L. Postman, The psychology of rumor. 1947.

[26] Jiarong Li, Haijun Jiang, Zhiyong Yu and ChengHu, "Dynamical analysis of rumor spreading model in homogeneous complex networks," *Applied Mathematics and Computation*, vol. (359), no. (15), pp. (374-385), 2019.

[27] Xiaobin Rui, Fanrong Meng, Zhixiao Wang, Guan Yuan and Changjiang Du, "SPIR: The potential spreaders involved SIR model for information diffusion in social networks," *Physica A: Statistical Mechanics and its Applications*, vol. (506), no. (15), pp. (254-269), 2018.

[28] Marcelo Kuperman and Guillermo Abramson, "Small World Effect in an Epidemiological Model," *Physical Review Letters*, vol. (86), no. (13), Article ID 2909, 2001.

[29] Adil Imad Eddine Hosni, Kan Li and Sadique Ahmad, "Analysis of the impact of online social networks addiction on the propagation of rumors," *Physica A: Statistical Mechanics and its Applications*, vol. (542), no. (15), Article ID 123456, 2020.

[30] Dongmei Fan, Guo-Ping Jiang, Yu-Rong Song and Yin-Wei Li, "Novel fake news spreading model with similarity on PSO-based networks," *Physica A: Statistical Mechanics and its Applications*, vol. (549), no. (1), Article ID 124319, 2020.

[31] Chun-Yan Sang and Shi-Gen Liao, "Modeling and simulation of information dissemination model considering user's awareness behavior in mobile social networks," *Physica A: Statistical Mechanics and its Applications*, vol. (537), no. (1), Article ID 122639, 2020.